3 April 2002                                                                                                                  1# GPM Draft Science Implementation Plan
# Ground Validation Chapter

S. Yuter, R. Houze, V. Chandrasekar,
E. Foufoula-Georgiou, M. Hagen, R. Johnson,
D. Kingsmill, R. Lawrence, F. Marks,
S. Rutledge, and J. Weinman.

Draft 1.2, 3 April 2002

material below assumes the following acronyms and symbols are defined earlier in the text

GPM – Global Precipitation Measurement
IFOV – instrument field of view
DPR – dual frequency precipitation radar
GMI – GPM Microwave Imager
TRMM – Tropical Rainfall Measuring Mission
$q_c$ – cloud water mixing ratio
$q_r$ – rain water mixing ratio
$q_i$ – ice mixing ratio
$q_s$ – snow mixing ratio
$q_a$ – aggregate mixing ratio
$q_g$ – graupel mixing ratio
$q_h$ – hail mixing ratio

# 1. Introduction

The validation of GPM satellite precipitation products is important for their credibility and utility within the larger community. The validation foci are primarily driven by the requirements of the GPM satellite algorithm developers and product customers. GPM product customer applications include data assimilation into atmospheric and hydrological models, climate diagnostics, and basic research into precipitation mechanisms and structure. The specific validation needs vary among the different algorithm developer and customer groups. The GV program will be designed to meet these diverse needs using a combination of near real-time routine products and focused surface-based and airborne observational data sets.

## 2. Objectives

The overall goals of GPM Ground Validation (GV) are as follows:

I. <u>Diagnosis</u> to ascertain the causes of errors within satellite products

II. <u>Improvement</u> of satellite products by refinement of physical assumptions in the satellite algorithms, underlying cloud resolving models, and underlying radiative transfer calculations.

III. <u>Evaluation</u> to estimate the quality of the satellite precipitation products in terms of systematic error and random error.

The validation of each of the GPM satellite precipitation product will need to address the following general objectives:

a) Determination of minimum detectable surface precipitation rate.
b) Classification of precipitation in (x,z) and (x,y) dimensions into hydrometeor categories such as rain, snow, mixed[1], and graupel/hail.
c) Classification of the three-dimensional precipitation structure
d) Spatial pattern of surface precipitation intensity.
e) Quantitative estimate of surface precipitation rate.
f) Description of errors associated with each of the above items (a-e).

The ground validation objectives have many similarities between the GPM active radar and passive microwave products. These areas of overlap will be discussed first in Section 2.1 and then instrument/algorithm specific validation objectives will be discussed in Section 2.2. Section 3 will discuss the approach to meeting these objectives in terms of components the GV program and the priorities for different types and locations of validation measurements.

---

[1] "Mixed" in this chapter refers to a mixture of ice, partially melted ice, and rain hydrometeors such as is found in the melting layer or in regions of freezing rain at the surface.

## *2.1 General objectives for validation of all GPM precipitation products*

### 2.1.1 Determination of minimum detectable surface precipitation rate

GPM satellite instruments will not have sufficient sensitivity and spatial resolution to detect the full range of precipitation rates measurable by surface-based instruments. For the GPM DPR on the core satellite, the minimum detectable threshold will be determined by the sensitivity of the radar and the IFOV. For the passive microwave sensors on the core and constellation satellites, the minimum detectable threshold will be determined by the exact frequencies employed, instrument sensitivity, IFOV, and the ability of the algorithms to isolate the precipitation signal from the background. For both the DPR and the passive microwave instruments, the translation of the sensors' minimum detectable signal into a minimum surface precipitation rate by hydrometeor type has several sources of complexity, and its determination will benefit from comparison to surface-based observations. Among the issues that will need to be addressed are the utility of GPM for precipitation estimation in snow and in stratocumulus drizzle, and the detection of non-precipitating cloud over the ocean by the core satellite sensors so that cloudless pixels can be identified for DPR calibration purposes.

### 2.1.2 Classification of precipitation into hydrometeor categories in (x,z) and (x,y) dimensions

At passive microwave and DPR frequencies, the radiative characteristics of basic hydrometeor categories such as rain, snow, mixed, and graupel/hail exhibit sufficiently large differences to be accounted for separately within the satellite algorithms. The estimation of the vertical distribution of hydrometeor categories (e.g. rain depth and ice depth) is an intermediary step and an output of the DPR algorithms and is a critical input to the passive microwave algorithms. Extra-tropical regions observed by GPM can have rain, snow, mixed, or graupel/hail surface precipitation types. Validation of hydrometeor categories in both vertical profiles (x,z) and horizontal maps (x,y) will be needed by GPM.

Within the deep convection of the tropics and summer midlatitude storms, the distribution with height of snow, mixed, and rain hydrometeor categories within a vertical column will usually be able to be inferred from information about the melting layer obtained by the DPR. Midlatitude winter convection can be deep (~10 km altitude) or shallow (< 3 km) depending on its position relative to the baroclinic front. Within shallow frontal precipitation over ocean, over land and within mountainous terrain, the differentiation among hydrometeor categories such as light rain (< 1 mm/hr), mixed, and snow will present a new set of challenges for the satellite algorithms.

### 2.1.3 Classification of precipitation structure

Classification of precipitation three-dimensional structure is important for a wide variety of applications related to GPM satellite products and to external users. An important GPM classification need is distinguishing among mesoscale convective systems, extratropical cyclones, and tropical cyclones as these storm structures have different energy sources and different relationships between their vertical profile of heating and the large-scale flow.

For validation of satellite latent heating products it is important to distinguish among subregions of the different classes of precipitating systems with significantly different vertical profiles of heating and/or with significantly different modes of propagation of heating. Within convective systems, the vertical profile of heating differs between convective and stratiform precipitation subregions and the heating is conveyed to the environment differently from the two subregions (Mapes and Houze 1995, Houze 1997). Methods developed for TRMM to classify convective and stratiform precipitation regions using 2D data and 3D data as input can likely be either adopted or adapted for GPM objectives. Research is needed to address whether the separation of convective/stratiform precipitation for latent heating applications is relevant to extratropical cyclones and to tropical cyclones or if different structural subcategories are needed.

## 2.1.4 Spatial pattern of surface precipitation intensity

The spatial variability of precipitation is dependent on scale and storm structure (Thiessen, 1911). In general as spatial scale at which the process is averaged increases, the standard deviation of rain rate decreases and the maximum rain rate decreases. Figure 1 illustrates an example of spatial scale versus the standard deviation of instantaneous rain rate for a snapshot of a radar-observed storm in Kansas (Tustison et al., 2001). Regional and seasonal differences in the exact relationship between spatial scale and the variability of precipitation are related to differences in precipitation structure. The surface-based observation, characterization, and comparison with satellite products of the spatial patterns of precipitation are important for GPM in two contexts: variability at scales smaller than the satellite IFOV and variability at scales larger than the satellite IFOV.

Precipitation exhibits spatial variability down to scales smaller than the smallest GPM IFOV of ~5 km x 5 km associated with the DPR. For electromagnetic sensors such as those on the GPM satellites, the received energy is a function of the incident energy and the number, size, and categories of hydrometeors within the volume of atmosphere encompassed by the IFOV but is largely independent[2] of the spatial distribution of the hydrometeors within the IFOV. The "beam-filling problem" relates to an assumption made for mathematical convenience that the hydrometeors are uniformly distributed in the volume. The uncertainties in precipitation estimation associated with beam filling may be able to be modeled (e.g. Wang 1996, Harris et al. 2002). Observations will be needed as input to the development of beam filling correction methodologies to address

---

[2] Independence to the spatial distribution assumes that the hydrometeors do not touch or block one another.

the spatial variability of precipitation from the scale of the 10.7 GHz GMI IFOV to a few 100 m.

Precipitation features larger than the satellite IFOVs can be compared to ground-based observations using statistics based on the absolute as well as the relative rainfall intensity (e.g. normalized by mean intensity). Evaluation of the location and spatial structure of relative precipitation intensity can provide important diagnostic and error characterization information on GPM satellite products and the underlying CRM and RTC models (Zepeda-Arce et al. 2000). These analyses do not require ground validation observations as rigorous as those associated with the evaluation of quantitative precipitation estimates. Objective comparison of the spatial patterns of precipitation intensity at appropriate scales is a logical preliminary step to quantitative precipitation evaluation since if the heavier versus lighter rain regions within the storm are not in the correct locations then the quantitative estimates will not be credible. Objective evaluation of the relative pattern of precipitation intensity can also provide information about the bias related to saturation of the satellite sensors and the relative precision of satellite-inferred rain rates through the full range of rain rates. Facilities such as networks of operational weather radars and rain gauges within several nations could provide information suitable for evaluation of relative precipitation patterns. Routine evaluation of the spatial pattern of precipitation intensity will be particularly informative over land where passive microwave algorithms are usually limited to use of the characteristics of the overlying ice to infer information about the rain layer.

### 2.1.5 Quantitative precipitation estimation

To be useful in the context of evaluation of satellite-derived quantitative precipitation estimates, the systematic and random errors in surface-based precipitation estimates need to be smaller than the systematic and random errors in the satellite precipitation products. This requirement is difficult to achieve in practice as surface-based observations must be performed and maintained under rigorous (and costly to implement) standards of quality control (e.g. Joss et al. 1998), the systematic and random errors must be robustly characterized for the entire range of rain rates, and the measurements must be accurately mapped into rain rates at spatial scales appropriate for comparison to the satellite products taking into account biases associated with the scale-dependent variability of precipitation (Tustison et al., 2002). Despite the practical difficulties, costs, and complexities, GPM ground validation will not be complete without the objective comparison of the pattern of absolute precipitation intensity within the area encompassed by at least one surface-based observation site. Among the issues that will need to be addressed under this topic are the saturation of satellite-estimated precipitation rate at high rain rates and the absolute precision of satellite-inferred rain rates through the full range of observed rain rates.

### 2.1.6 Description of errors

It is vital that the errors associated with observed and inferred variables in GPM products be adequately described so that their associated uncertainties can be appropriately accounted for in error modeling. Observational data, derived estimates, derived constants in functional relations, and GV products will be considered incomplete without an accompanying error description. A common method of error description suitable for many spatially varying geophysical variables is to report error characteristics in terms in three basic components:

- Estimates of the mean value and the standard deviation of the systematic error within a pixel.
- Estimate of the standard deviation of the random error within a pixel.
- Estimates of functions describing the spatial correlation of systematic errors with distance and random errors with distance (x,z or x,y) between pixels.

Systematic errors are reproducible discrepancies between a variable and its "true" value (Bevington and Robinson 1992). Accuracy describes how close an estimate is to the true value and depends on how well systematic errors can be minimized or compensated for. Random errors are fluctuations in observations that vary from measurement to measurement (Bevington and Robinson 1992). By definition random errors are assumed to have a mean of zero. Precision is a measure of the reproducibility of the result and is dependent on how well random error can be minimized. The specific method of error description for a particular variable or product should satisfy the joint constraints of being appropriate to the variable and its measurement method and of being suitable for ingest into applications utilizing that variable or product.

Error propagation translates estimates of uncertainties in observed values into a total estimate of uncertainty in a derived variable. For example, for derived variable α, which is a function of observed variables $u$ and $v$ ($\alpha=f(u,v)$), the variance of α ($\sigma_\alpha^2$) can be approximated as:

$$\sigma_\alpha^2 \cong \sigma_u^2 \left(\frac{\partial \alpha}{\partial u}\right)^2 + \sigma_v^2 \left(\frac{\partial \alpha}{\partial v}\right)^2 + 2\sigma_{uv}^2 \left(\frac{\partial \alpha}{\partial u}\right)\left(\frac{\partial \alpha}{\partial v}\right) \qquad [1]$$

Where the variances of $u$ ($\sigma_u^2$) and $v$ ($\sigma_v^2$) are given by

$$\sigma_u^2 = \lim_{N \to \infty} \left[\frac{1}{N} \sum (u_i - \bar{u}_i)^2\right] \text{ and } \sigma_v^2 = \lim_{N \to \infty} \left[\frac{1}{N} \sum (v_i - \bar{v}_i)^2\right] \qquad [2]$$

and the covariance $\sigma_{uv}^2$ is given by

$$\sigma_{uv}^2 = \lim_{N \to \infty} \left[\frac{1}{N} \sum (u_i - \bar{u}_i)(v_i - \bar{v})\right] \qquad [3]$$

Covariances are non-zero when variables are correlated with one another. Independent variables are by definition not correlated and hence are preferred for error modeling when possible. Of particular interest to data assimilation are the spatial covariances of systematic error and random error. Spatial covariance for a variable $u$ with distance $\Delta x$ is given by:

$$\sigma^2_{u(\Delta x)} = \lim_{N \to \infty}\left[\frac{1}{N}\sum (u_x - \bar{u})(u_{x+\Delta x} - \bar{u})\right] \quad [4]$$

Error modeling is based on values for the variances and covariances of input variables and a functional description of the relation between the input variables and the derived dependent variable. A goal of GV error description is to provide input to error modeling in terms of values for variances and covariances corresponding to a selected set of observed and derived variables associated with the GPM products. This goal also implies error modeling internal to GV. For example, an error propagation equation for the derivation of surface rain rate including all sources of uncertainty will be needed to derive the variance of rain rate in the GV rainfall products. The selection of the subset of GV error description variables will be done jointly with the satellite algorithm and application teams.

## 2.2 Instrument/Algorithm Specific Needs

### 2.2.1 Dual Frequency Radar

The GPM core satellite DPR will yield high vertical resolution profiles of radar reflectivity ($Z$) at 13.6 GHZ and 35 GHz (wavelengths of 2.3 cm and 0.85 cm). In comparison to the 5 cm and 10 cm wavelength operational surface-based radars used by national weather services worldwide, the DPR will experience more attenuation within precipitation and its measured reflectivities will have a larger contribution from non-Rayleigh scattering (Meneghini and Kozu 1990).

The dual frequency design of the DPR will utilize the differential attenuation of the returned signals to infer information about the bulk characteristics of the particle size distribution (e.g. $D_o$, the diameter that divides the rain water content ($q_r$) into two equal parts) and hydrometeor category (e.g. rain, snow, mixed, wet graupel/hail). Particularly at 35 GHz, cloud drops may contribute to integrated attenuation of precipitating clouds as a function of cloud depth and possibly the characteristics cloud drop spectra (e.g. marine versus continental clouds).

The quantification of the attenuation of radar reflectivity with height is an important component in the estimation of near-surface precipitation. Methods to correct for attenuation in space-borne radar include backscatter, single-wavelength surface reference, dual-wavelength surface reference, dual-wavelength, and mirror-image (Meneghini and Kozu 1990). These methods have various strengths and weaknesses and it is likely that some combination of them will be implemented for the GPM DPR algorithm. Land versus ocean surfaces present different challenges for the surface reference methods. Over land, the scattering cross-section of the surface ($\sigma°$) varies with incidence angle, terrain, and ground cover (Meneghini and Kozu 1990). Over ocean, $\sigma°$ varies with incidence angle, wind speed, and precipitation rate (Contreras et al. 2002).

Validation needs associated with the refinement of physical assumptions within the DPR algorithms fall into three main categories:

- Information on the instantaneous spatial distributions (x,z) and (x,y) (e.g. multiscale variability) and climatological variability (e.g. probability distributions) of bulk water content ($q_c$, $q_r$, $q_i$, $q_s$, $q_a$, $q_g$, $q_h$) and of particle size distribution (N(D)) by hydrometeor type.

- Measurements of $\sigma°$ over water within precipitation under varying rain rate and wind conditions at DPR frequencies and incidence angles.

- Measurements of $\sigma°$ over snow covered surfaces to characterize variability within 5 km footprints in mountainous terrain.

## 2.2.2 Passive Microwave

The physics of the passive microwave algorithms is largely contained in the cloud resolving models (CRM) and radiative transfer calculations (RTC) that underlie them. The output of a model is dependent on the quality of the model's input as well as the physical assumptions embedded within it. Hence refinement of physical assumptions within the passive microwave algorithms is directly related to the quality of the input and the refinement of physical assumptions within the CRM and RTC.

**2.2.2.1 Cloud resolving models**

A cloud-resolving model can be defined as a model in which the cloud physics is explicitly defined and no cumulus parameterizations are required. Increases in the speed of computer processing have lead to a blurring of the distinction between cloud models and regional models that is likely to continue. Environmental observations used to initialize cloud and regional models include: high resolution boundary layer and coarser resolution tropospheric profiles of $u$, $v$, T, and RH, SST, soil moisture, snow cover, vegetation ground cover, and surface short-wave and long-wave fluxes. In additional to these environmental observations, data assimilated into regional models often includes the spatial pattern of surface meteorological variables (winds, T, RH, pressure), and the three-dimensional wind pattern (Xue et al., 1995). The time and spatial resolution of the wind and thermodynamic observations needed to initialize models varies and will have to be tailored to each domain of interest. Model simulations of storms developing in relatively uniform environments or with simple topographically fixed forcing such as continental mesoscale convective systems and non-frontal orographic precipitation can be initialized with a single upper air sounding. For storms developing in nonuniform conditions or associated with complex forcing, the time varying, three-dimensional structure of the environment must be adequately described in the model initialization data and networks of research-quality soundings at high time resolution will be needed.

The physics within cloud and regional models usually yields physically consistent output but not always realistic output. An important component of model refinement is objective comparison of model output to independent observations (i.e. observations that were not used in the initialization or data assimilation steps) obtained by both ground validation and satellite instruments. Methodologies similar to those described in Section 2.1 can be employed to diagnose, improve, and evaluate the 2D aspects of model output. Additionally, methods suitable for statistical comparison of three-dimensional fields in time will also be needed (e.g. Yuter and Houze 1995). Refinement of physical assumptions within cloud and regional models will need:

- Microphysical information on the three dimensional (x,y,z) instantaneous spatial distributions (e.g. multiscale variability) and climatological variability (e.g. probability distributions) of bulk water content ($q_c, q_r, q_i, q_s, q_a, q_g, q_h$) and of particle size distributions (N(D)) by hydrometeor type within precipitating regions and within convective and stratiform subregions of the precipitating systems.

- Statistical information on precipitating cloud systems (e.g. population statistics, height distributions, and areal distributions.)

- Information on the three-dimensional wind and thermodynamic fields at high time and spatial resolutions for the region covered by the model.

**2.2.2.2 Radiative transfer calculations**

Cloud models produce many of the inputs to the RTC and thus the accuracy and precision of the RTC is directly related to how well the cloud model performs. Additionally, the boundary conditions of the RTC require knowledge of the surface emissivity. The land emissivities are especially complex as they can depend on antecedent snowfall, soil moisture, and frozen mud effects as well as the extent to which the ground is covered by rock outcroppings and trees. Land surface emissivity will need to be documented and updated with time. RTC are also sensitive to the treatment of ocean surface reflection (Smith et al. 2002) and the assumptions regarding the radiative properties of aspherical ice crystals (Wu and Weinman 1984). Surface-based and airborne observations necessary for refinement of physical assumptions in the RTC include surface roughness at GMI frequencies, mixture of surface cover (e.g. outcroppings of rocks and trees over snow), and the density and shapes of ice crystals. Of particular importance are observations that would improve modeling of melting snowflakes and freezing drops in the mixed phase region. Water vapor profiling within precipitating pixels will be especially important to estimation of precipitation over land at frequencies > 100 GHz. Low altitude water vapor screens the surface and high altitude water affects the sensitivity of those brightness temperatures to snow scattering.

## 2.2.3 Latent Heating Profile Algorithm

A latent heating product has been proposed for GPM that would provide information on the vertical distribution of heating within convective and stratiform precipitation. *Heating cannot be observed directly*, but rather must be inferred indirectly from satellite-derived estimates of instantaneous hydrometeor distributions. CRMs have been used to relate model output heating to quantities measured by the satellite instruments. GPM-product latent heating profiles will be derived from input from the CRMs, so the validation needs are similar to those in Section 2.2.2.1. Detailed and accurate descriptions of the large-scale wind and thermodynamic forcing are critical to high quality CRM simulations of heating. Wind and thermodynamic profile observations including stations surrounding the model domain and within the domain are needed to characterize the large scale forcing for CRMs. Independent validation of the heating profiles in the tropics will benefit from budget studies based on high time resolution networks of upper air sounding data and high-resolution three-dimensional wind fields (e.g. from dual-Doppler radar) that can be partitioned into regions containing convective and stratiform precipitation.

In the tropics, the vertical profile of heating on the large scale is determined almost entirely by the latent heat released in convective systems (Houze 1989, Mapes and Houze 1995), which draw their energy from the unstable thermodynamic stratification. Assessing the vertical distribution of latent heat release is fundamental to tropical dynamics. A key aspect of the GPM latent heating products will be the maximum level of heating because it is the vertical gradient of the latent heating that affects the large-scale balanced circulation (as seen in the large-scale potential vorticity equation, Haynes and MacIntyre 1987). The vertical profile of heating differs between the convective and stratiform subregions of convective systems (Houze 1982, Johnson 1984), and the heating is conveyed to the environment differently from the two subregions (Mapes and Houze 1995, Houze 1997). Horizontal divergence and vertical velocity estimated from 3D wind fields can used to verify aspects of the latent heating profiles (e.g. Mapes and Houze 1995). Most attention to the problem has been in connection with tropical convection. However, midlatitude convective systems also divide themselves into convective and stratiform subregions, which behave similarly to their tropical counterparts (e. g. Braun and Houze 1996).

In midlatitudes, much of the precipitation comes from extratropical cyclones, for which large-scale baroclinicity is the driving energy source and for which latent heating profiles relate differently and more passively to the large-scale dynamics than in the case of deep convection. Since the dynamics of extratropical cyclones is fundamentally different from deep convection, the TRMM-like convective/stratiform separation of precipitation patterns may not be relevant to extratropical cyclones. There will need to be a way to distinguish the extratropical cyclonic precipitation systems from the convective systems. In addition, tropical cyclones account for significant precipitation in both the tropics and mid-latitudes. Although tropical cyclones derive their energy from the tropical oceanic boundary layer, the TRMM-like convective/stratiform separation of precipitation patterns may yet be relevant to these storms since they exhibit precipitation structure with convective and stratiform regions similar to those seen in mesoscale convective systems (Marks and Houze 1987, Houze 1993). Nonetheless, since tropical cyclones are neither like deep convection nor like extratropical cyclones, they too need to be dealt with separately in terms of latent heating profiles.

## 2.2.4 Data Assimilation

The goal of data assimilation is to produce a regular, physically consistent, four dimensional representation of the atmosphere from a heterogeneous collection of in situ and remote sensing instruments that sample imperfectly and irregularly in space and time (Schlatter et al. 1999). Without an accurate description of the error characteristics, assimilation of an observed or inferred variable into a numerical model can do more harm than good.

### 2.2.4.1 Global Models

Assimilation of precipitation-related data into global models uses a special subcategory of error characterization that constrains the description of the errors be unbiased, normally distributed, and homogeneous in space and time. In contrast to more continuous fields such as pressure, errors in discontinuous fields such as rainfall rate are not usually normally distributed and homogeneous in space and time. A transformation will be needed to relate the errors in the GPM precipitation products into a normally distributed and homogenous form. Information on the nature of the errors in terms of independent sources and each source's relative contribution to total error is more important than exact numerical estimates as once a mathematical description of the errors is available the numerical error estimation can be done within the data assimilation process. Sources of uncertainty include: sensor measurement errors, uncertainty associated with multi-scale variability (e.g. representativeness error), errors in the derivation of the inferred variable from the observed variable, and errors in physical assumptions in the CRM and RTC. Separate error descriptions are desired for regions with significantly different heating profiles such as for convective and stratiform precipitation within convective systems.

### 2.2.4.2 Regional Atmospheric and Hydrological models

Forecast applications of GPM will involve the routine ingest of GPM information by regional and hydrological models. Utilization of real-time GPM satellite products by regional models of coastal regions in particular has a strong potential to yield measurable benefits to forecast accuracy. Algorithm design, data formats and the description of errors suitable for assimilation in regional and hydrological models may have different requirements than global models. Regional models will also play an important role in the refinement of physical assumptions associated with GPM. GV activities can serve as a testbed for development of tailored input data streams, error descriptions, and data formats to serve regional and hydrological model applications.

Table 1. Preliminary mapping of measurements to validation application categories based on group discussion at GPM GV workshop Feb 2002.

| | Observations | | | | | | | | Algorithms | | | | | |
|---|---|---|---|---|---|---|---|---|---|---|---|---|---|---|
| | Z(14,35) | Tb10 | Tb19 | Tb21 | Tb37 | Tb85 | Tb150 | Tb183 | Rs | E(Rs) | LH(z) | E(LH) | DSD(Z) | E(DSD) |
| *Remote sensing* | | | | | | | | | | | | | | |
| Passive MW radiom. | 10 | 10 | 10 | 10 | 10 | 5 | 5 | 5 | 10 | 10 | 8 | 8 | 0 | 0 |
| Non-att. VP radar | 10 | 8 | 8 | 8 | 8 | 8 | 8 | 8 | 8 | 8 | 10 | 10 | 8 | 8 |
| Att. VP radar | 10 | 8 | 8 | 8 | 8 | 8 | 8 | 8 | 8 | 8 | 8 | 8 | 9 | 9 |
| Scanning Dop Radar | 10 | 10 | 10 | 10 | 8 | 7 | 6 | 5 | 10 | 10 | 10 | 10 | 8 | 8 |
| attenuation measurement | 4 | 0 | 0 | 0 | 0 | 0 | 0 | 0 | 4 | 4 | 0 | 0 | 0 | 0 |
| Scanning pol. Radar | 10 | 10 | 10 | 10 | 9 | 8 | 7 | 6 | 10 | 10 | 10 | 10 | 10 | 10 |
| *Surface precipitation measurement* | | | | | | | | | | | | | | |
| Disdrometer net | 7 | 0 | 0 | 0 | 0 | 0 | 0 | 0 | 10 | 10 | 8 | 8 | 10 | 10 |
| hydrometeor type | 7 | 2 | 2 | 2 | 2 | 2 | 2 | 2 | 10 | 10 | 8 | 8 | 10 | 10 |
| snow gauge | 5 | 0 | 0 | 0 | 0 | 5 | 5 | 5 | 10 | 10 | 8 | 8 | 5 | 5 |
| rain gauge net | 5 | 0 | 0 | 0 | 0 | 0 | 0 | 0 | 10 | 10 | 8 | 8 | 5 | 5 |
| *Environment observations* | | | | | | | | | | | | | | |
| Boundary layer profile of u,v,T,q. | 10 | 10 | 10 | 10 | 10 | 10 | 10 | 10 | 10 | 10 | 10 | 10 | 3 | 3 |
| tropospheric u,v,T,q profiles | 10 | 10 | 10 | 10 | 10 | 10 | 10 | 10 | 10 | 10 | 10 | 10 | 3 | 3 |
| SST | 3 | 7 | 7 | 7 | 7 | 7 | 7 | 7 | 7 | 7 | 7 | 7 | 3 | 3 |
| sfc sh & lh flux | 3 | 8 | 8 | 8 | 8 | 8 | 8 | 8 | 8 | 8 | 10 | 10 | 3 | 3 |
| ocean roughness | 10 | 5 | 5 | 5 | 5 | 5 | 5 | 5 | 10 | 10 | 0 | 0 | 0 | 0 |
| *Microphysical variables* | | | | | | | | | | | | | | |
| VPH(qc,qi, qr, qs, qg, qh) | 10 | 10 | 10 | 10 | 10 | 10 | 10 | 10 | 10 | 10 | 10 | 10 | 10 | 10 |
| N(D) for qc,qi, qr, qs, qg, qh | 10 | 10 | 10 | 10 | 10 | 10 | 10 | 10 | 10 | 10 | 10 | 10 | 10 | 10 |
| hydro shape, ρ, Vt | 10 | 0 | 0 | 0 | 8 | 10 | 10 | 10 | 10 | 10 | 0 | 0 | 8 | 8 |
| melting rate | 8 | 5 | 8 | 8 | 8 | 8 | 8 | 8 | 8 | 8 | 10 | 10 | 0 | 0 |
| hydro evap rate | 8 | 8 | 8 | 8 | 8 | 5 | 5 | 5 | 8 | 8 | 10 | 10 | 7 | 7 |
| riming rate | 0 | 0 | 0 | 0 | 0 | 0 | 0 | 0 | 8 | 8 | 10 | 10 | 10 | 10 |
| coalescence rate | 0 | 0 | 0 | 0 | 0 | 0 | 0 | 0 | 8 | 8 | 0 | 0 | 10 | 10 |
| deposition rate | 0 | 0 | 0 | 0 | 0 | 0 | 0 | 0 | 8 | 8 | 10 | 10 | 10 | 10 |
| condensation rate | 1 | 0 | 0 | 0 | 0 | 0 | 0 | 0 | 8 | 8 | 10 | 10 | 10 | 10 |
| aggregation rate | 1 | 0 | 0 | 0 | 0 | 0 | 0 | 0 | 8 | 8 | 0 | 0 | 10 | 10 |
| drop breakup rate | 1 | 0 | 0 | 0 | 0 | 0 | 0 | 0 | 5 | 5 | 0 | 0 | 5 | 5 |
| cloud base | 5 | 5 | 5 | 5 | 5 | 5 | 5 | 5 | 5 | 5 | 5 | 5 | 5 | 5 |
| CCN | 5 | 5 | 5 | 5 | 5 | 5 | 5 | 5 | 6 | 6 | 6 | 6 | 8 | 8 |
| lightning | 5 | 5 | 5 | 5 | 5 | 5 | 5 | 5 | 5 | 5 | 5 | 5 | 5 | 5 |
| *Hydrological variables* | | | | | | | | | | | | | | |
| runoff | 0 | 0 | 0 | 0 | 0 | 7 | 10 | 10 | 10 | 7 | 0 | 0 | 0 | 0 |
| snowpack | 0 | 0 | 0 | 0 | 0 | 0 | 10 | 10 | 0 | 0 | 0 | 0 | 0 | 0 |
| soil moisture (in situ) | 10 | 10 | 8 | 5 | 1 | 0 | 0 | 0 | 10 | 7 | 0 | 0 | 0 | 0 |
| surface meteor. Measure (in situ) | 0 | 0 | 0 | 0 | 0 | 0 | 0 | 0 | 5 | 5 | 0 | 0 | 0 | 0 |

## *2.3 Preliminary Mapping of Measurement Types to Validation Applications*

At the 1st GPM GV workshop in Feb 2002, a preliminary mapping of measurements to validation application categories was discussed. The output of that discussion is shown in Table 1, which uses a relative scale from 0 to 10, indicating "not important" (0) to "very important" (10). Validation application categories are described in terms of satellite sensors and generic product types. Measurement types and instruments are indicated with generic descriptions to leave open exploitation of future instruments and methodologies. The table represents a distillation of algorithm developer and product customer needs gleaned from the discussions at the workshop and is a summary of the needs described and implied in the sections above.

## 3. Approach

The GPM GV program has two primary programmatic components:

- Routine Product Site (RPS) – Multi-year observation program overlapping with GPM satellite operational period that produces near real-time routine GV products on a regular schedule.

- Focused Measurement Program (FMP) – Observational program that produces data and derived products addressing GPM algorithm and model validation needs but does not directly produce routine GV products on a regular schedule. FMPs include field programs involving aircraft and focused measurement programs without aircraft. Timing and location for measurement programs are optimized on science and cost effectiveness. Some may be co-located with RPS.

The following subsections provide further details on the RPS and FMPs.

## *3.1 Routine Product Sites*

Several RPS in different geographic regions will be needed since error characteristics differ where different portions to the GPM satellite algorithms are used. Table 2 summarizes the three basic satellite algorithm regimes in terms of the varying utilization of emission and scattering channels and the ocean surface reference constraint.

Table 2. Algorithm regimes with different error characteristics.

| Algorithm Regimes | Passive Microwave Algorithms | Radar Algorithm Constraints |
|---|---|---|
| Open Ocean | Emission and Scattering | Ocean surface reference |
| Continent and Mountains | Primarily Scattering for frequencies < 100 GHz. Emission and scattering for frequencies > 100 GHz. | W/o ocean surface reference |
| Near Coasts | Primarily Scattering for frequencies < 100 GHz. Emission and scattering for frequencies > 100 GHz. | Ocean surface reference |

Analysis is needed to determine if the differences between the error characteristics of tropical versus midlatitude precipitation are sufficient to warrant separate RPS in tropical *and* midlatitude open ocean, continental, and coastal sites.

The RPS will focus on two related but distinct activities:

- GV Local Products – research, development, and near real-time implementation of a suite of products based on routine data collected at the site to address diagnosis and evaluation of GPM satellite products and improvement of physical and scaling assumptions.

- GV Global Products – research and development of error description products for relevant subset of points (e.g. tropical ocean, mid-latitude continent) within each GPM satellite precipitation product.

GV Local Products will include products similar to the original TRMM GV products such as Cartesian volumes of observed variables like radar reflectivity, Cartesian maps of derived variables like convective and stratiform precipitation subregions and estimated surface rain rate, and statistical summaries of volumetric fields such as vertical profiles and CFADs (Yuter and Houze 1995). The product list will be expanded to include routine upper-air soundings in a standard format at all RPS and site-appropriate utilization of polarization diversity data, dual frequency data, and multiple Doppler data where available. The production of GV Local Products is the responsibility of each RPS. The Goddard DAAC will handle worldwide dissemination of the GV Local Products. To increase utilization of these local products by a wide range of users, the goal will be to have them processed and accessible online within 48 hours of data collection. Local products in combination with the RPS overpass subset of GPM satellite products will constitute a test bed data set supporting diagnosis, refinement, and evaluation activities and the development of the GV Global Product algorithm.

GV Global Products will be the output of a combined algorithm developed at several of the RPSs The GV Global Product algorithm data processing will be run at a facility at Goddard. Based on the geographic location of the satellite pixel, the GV Global Product algorithm will call the appropriate part of the combined algorithm associated with the relevant RPS algorithm regime. Inputs to the GV Global Product algorithm can include any of the products within the GPM data stream. GV Global Products will provide information on error descriptions (Sections 2.1.6 and 2.2.4).

The GV Global Products are intended to represent a different approach to error characterization than that provided by the GPM satellite algorithm developers. Error models are only as good as the quality of their inputs and their degree of completeness. The hardest errors to detect are the errors of omission. By providing an approach to error characterization based on a combination of the GV general objectives: detection, classification, spatial pattern, and quantitative estimation (Section 2), the GV Global Product may reveal errors of omission in the GPM satellite algorithm error characterizations which will aid in diagnosis and refinement of the satellite algorithms. Similarly, the GPM satellite algorithm error characterizations may reveal errors of omission in the GV Global Product algorithms. Having two approaches to error characterization on the same set of products will likely make the combined error characterization and the final GPM products better than only one approach could provide.

### 3.1.1 Requirements for RPS

Based on the discussion above, several minimum requirements for the RPS can be defined:

- <u>Scanning Doppler radar located within 125 km of a 12 hourly upper-sounding station.</u> A scanning Doppler radar is the centerpiece of the RPS providing observations of reflectivity and radial velocity and selected polarization diversity variables where appropriate. A 12 hourly operational upper air station within the coverage area of the radar products is the minimum requirement associated with initializing CRM which will be an integral component of GV activities.

- <u>GV Local Products are produced at least 6 months/year</u>. Most areas have wet and dry seasons. To justify the infrastructure expenses associated with the RPS, a minimum number of products months during the wet season are required. A 1-2 month period of scheduled down time for product production during the dry season when GV Local Products are not produced would provide a yearly window for maintenance and upgrades of site facilities.

- <u>Path to error characterization for global algorithm regime represented by RPS and associated FMPS.</u> The combination of the RPS and associated FMPs must provide sufficient information upon which to base a Global GV Product for the algorithm regime. Among the issues to be addressed are the variations of storm structures and surface precipitation types within the algorithm regime and elevation issues associated with applying measurements taken in elevated terrain to sea level.

To facilitate comparison of GV Local Products among sites, it would be useful to set minimum climatological precipitation criteria for rainfall frequency and accumulation associated with a month during which GV Local Products are produced. Frequency of precipitation is more important than accumulation since much of the research and development will be based on the subset of RPS data obtained during GPM satellite overpasses with precipitation.

## *3.2 Focused Measurement Programs*

Geography and climatology preclude addressing the full set of crucial GPM measurement objectives at the small number of RPS. FMPs fill in these gaps with remote sensing and in situ measurements. FMPs include long-term monitoring and field programs with aircraft. FMP's scientific goals are required to address one or both of the following:

- Refinement of specific physical and scaling assumptions within satellite algorithms, cloud models, and/or radiative transfer calculations.

- Initialization/data assimilation observations for cloud and regional modeling and/or radiative transfer calculations at climatologically important locations distinct from the RPS.

From a programmatic standpoint, any measurement that does not directly contribute to a GV Local Product is part of a FMP. Since these data are not part of the tightly regulated GPM data stream, the utility of the FMP data will be dependent on their timeliness and data format. To the degree possible, these measurements should be provided to the wider community in public domain, community data formats (e.g. ASCII text, hdf, netcdf, UF and DORADE for radar data, and CFPD and CMPD for aircraft microphysics data).

### 3.2.1 Candidate FMPs

At the 1$^{st}$ GPM GV Workshop in February 2002, several candidate FMPs were discussed (Table 3). Candidate pre-launch FMPs address GPM science issues associated with midlatitude and orographic precipitation that cannot be adequately addressed with TRMM satellite data, data from the TRMM Field Campaigns, or with data from other TRMM GV or AQUA activities. These pre-launch FMPs need to occur sufficiently prior to launch to have an impact on Day 1 GPM algorithms. The relative priority of post-launch FMPs will be determined based on analysis of GPM data and the degree other observations are needed to resolve relevant issues.

Table 3. Candidate FMPs from discussion at 1st GPM GV Workshop, February 2002.

| Timing | Target | Scientific Objective | Potential Location | Platforms | Notes |
|---|---|---|---|---|---|
| Pre-launch | Midlatitude open ocean baroclinic storms | Rain-mixed-snow transition for water flux and LH retrievals | Gulf of Alaska, North Atlantic | Multiple aircraft | Overlaps with coastal activities |
| | Midlatitude coastal mountains | GPM data assimilation issues for regional models, phys. assumptions for CRM, RTC in orog. precip. | U.S. West coast | Aircraft and surface-based obs in mtns and looking offshore | Overlaps with open ocean activities |
| | Midlatitude prairie | GPM data assimilation issues for regional models, phys. assumptions for CRM, RTC. Simple land boundary conditions for radar and radiometry. | Canadian prairie provinces or US upper midwest | Aircraft and surface based obs | Overlaps with coastal activities |
| Post-launch | Regions exhibiting significant discrepancies in est. precip among diverse sensors | Diagnose errors and refine physical assumptions in CRM and RTC to resolve discrepancies | East Pacific ITCZ | Ship | Build on PACS data sets |
| | Variability of oceanic precip structures distinct from RPS[3] | Refinement of physical assumptions in CRM and RTC to address range of storm structures | Indian Ocean Monsoon, South Pacific Convergence Zone | Ship, aircraft and land-based obs | Cooperation with other programs when possible. |
| | Hurricanes | Refinement of physical assumptions in satellite algorithms associated high R | Hurricane regions worldwide | Aircraft, possibly ship and land-based obs. | Cooperation with other programs is necessary. |

---

[3] Analysis of satellite data will indicate how well the GPM satellite by itself is discerning the structural differences among the regions and if there are sufficient issues not addressable by the satellite data to warrant other observations.

| | | | | | |
|---|---|---|---|---|---|
| | Tropical coastal mountains[3] | Refinement of physical assumptions in CRM and RTC for tropical orog. precip. | Tropical coastal mountains with copious rainfall | Aircraft and surface-based obs in mtns and looking offshore | Hurricane landfall a priority. |
| | Midlatitude prairie | Refinement of physical assumptions in CRM and RTC to address snow and light rain over land | Canadian prairie provinces or US upper midwest regions with copious snow. | Surface-based obs and aircraft | Important for estimation of surface snow and light rain rates over simple terrain. |

3 April 2002                                                                                                     20

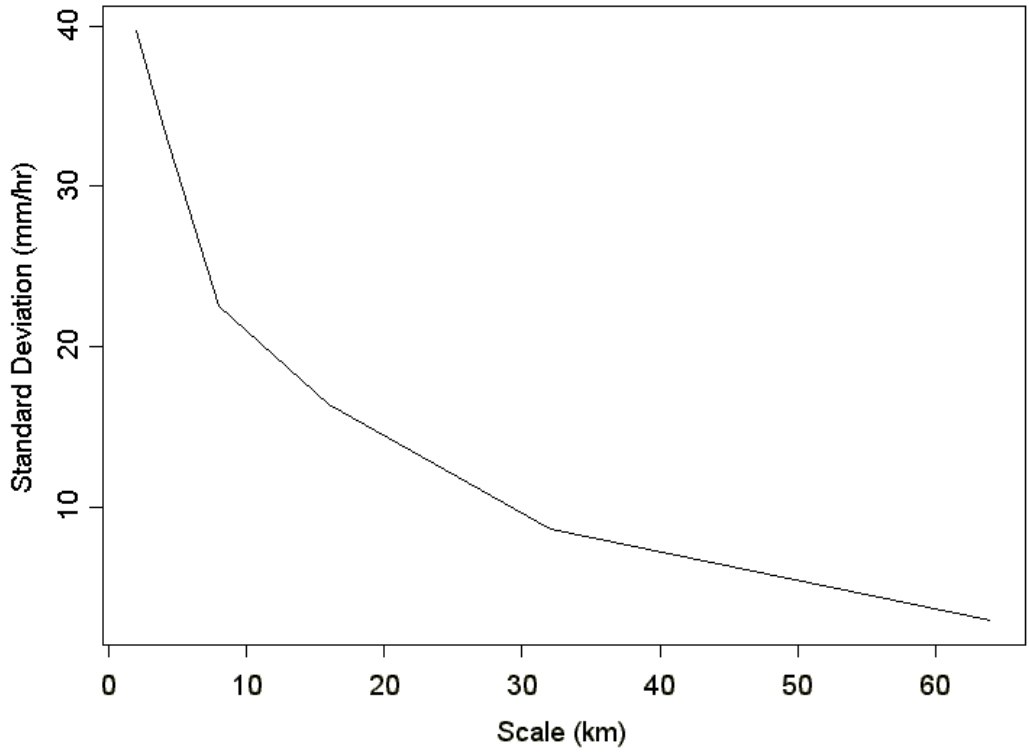

Figure 1. Standard deviation of nonzero precipitation as a function of grid scale for radar observed precipitation from the NEXRAD KICT at 18:47 UTC 17 August 1994. Adapted from Tustison et al. 2001.